\documentclass[aps,prb,twocolumn]{revtex4}
\usepackage{graphicx}
\usepackage{amssymb}
\usepackage{amsfonts}
\usepackage{amsmath}
\usepackage{bm}
\usepackage{graphicx}
\usepackage{color}

\renewcommand{\vec}[1]{\mathbf{#1}}
\newcommand{\cc}[1]{{#1}^\star}
\newcommand{\pd}{\ensuremath{p\text{-}d}}
\newcommand{\etal}{\textit{et al.}}
\newcommand{\kp}{\ensuremath{k.p}}
\DeclareMathOperator{\im}{Im}

\begin{document}

\title{Effect of inversion asymmetry on the intrinsic anomalous Hall effect\\ in ferromagnetic (Ga,Mn)As}

\author{Agnieszka Werpachowska}
\affiliation{Institute of Physics, Polish Academy of Sciences, al.~Lotnik\'ow 32/46, PL-02-668 Warszawa, Poland}

\author{Tomasz Dietl}
\affiliation{Institute of Physics, Polish Academy of Sciences, al.~Lotnik\'ow 32/46, PL-02-668 Warszawa, Poland}
\affiliation{Institute of Theoretical Physics, University of Warsaw, PL-00-681 Warszawa, Poland}

\begin{abstract}
The relativistic nature of the electron motion underlies the intrinsic part of the anomalous Hall effect, believed to dominate in ferromagnetic (Ga,Mn)As. In this paper, we concentrate on the crystal band structure as an important facet to the description of this phenomenon. Using different $k.p$ and tight-binding computational schemes, we capture the strong effect of the bulk inversion asymmetry on the Berry curvature and the anomalous Hall conductivity. At the same time, we find it not to affect other important characteristics of (Ga,Mn)As, namely the Curie temperature and uniaxial anisotropy fields. Our results extend the established theories of the anomalous Hall effect in ferromagnetic semiconductors and shed new light on its puzzling nature.
\end{abstract}

\maketitle

\section{Introduction}
The anomalous Hall effect was first observed in ferromagnets by Hall himself.\cite{hall:1880} Next to the usual Lorentz term, a voltage proportional to magnetization appeared -- much smaller, but still too large to be explained by an internal magnetic field. Over the following years, this ostensibly plain dependence was to unveil the whole cornucopia of physical phenomena, all arising from the relativistic coupling of the charge and spin current.\cite{dyakonov:2007,Nagaosa:2009a} A class of them is related to the spin asymmetry of carrier scattering, \textit{viz}.\ the skew-scattering\cite{smit:1955,mott:1965,dyakonov:1971} and the side jump process.\cite{berger:1970,berger:1972,nozieres:1973} Also higher-order effects in the scattering amplitude were predicted.\cite{mott:1965,dyakonov:1984,dyakonov:2007,abanin:2009} However, it is the ``intrinsic'' mechanism, first proposed by Karplus and Luttinger,\cite{karplus:1954} that is believed to play a key role in the Hall effect of ferromagnetic semiconductors.\cite{matl:1998,taguchi:2001,jungwirth:2002,Dietl:2003c,lee:2004a,lee:2004b} Quite unusually, it does not concern the changing of the occupations of Bloch bands by scattering on impurities. Rather, it results from the interband coherence caused by the universal tendency of physical systems to progressively increase the indeterminacy of their state. Hence, the topological theory of the Berry phase\cite{berry:1984} was found to provide an ample description.\cite{sundaram:1999,haldane:2004}

The topological model of the intrinsic anomalous Hall effect describes the linear response of the carrier Bloch function $\psi$ to the applied electric field $\vec E$. It consists in the drift of the $\vec k$ vector in the reciprocal space, during which $\psi(\vec k)$ acquires a geometrical phase factor in addition to the dynamical one. The Berry phase in the first factor can be expressed as the action of the vector potential, with its curl called the Berry curvature, $\boldsymbol\Omega(\vec k)$. The latter is a well-defined gauge-invariant quantity often pictured as a non-homogeneous magnetic field living in $k$-space. It produces a local equivalent of the Lorentz force, the so-called anomalous velocity term, $-e \vec E \times \boldsymbol\Omega$, in the semiclassical equations of motion.\cite{karplus:1954,sundaram:1999} This term contributes to the stationary part of the Boltzmann transport equation (hence, it does not depend on the transport relaxation time), producing a dissipationless current transverse to $\vec E$. The anomalous Hall conductivity of this current is proportional to the ensemble average of the carrier Berry curvature, $\langle \boldsymbol\Omega \rangle$. Since $\boldsymbol\Omega$ depicts the changes of the spin polarization during the carrier transport by the electric field, which are caused by the spin-orbit coupling, it changes sign under time-reversal symmetry. Thus, to obtain a finite value of $\langle \boldsymbol\Omega \rangle$, this symmetry of the system must be broken.

The above semiclassical approach, taking into account the complete geometrical Bloch state description, leads to an intuitive picture of the origin and mechanism of the intrinsic anomalous Hall effect in ferromagnetic semiconductors. It applies to the weak scattering regime, where it was proven to be formally equivalent to more systematic quantum-mechanical techniques.\cite{sinitsyn:2008} In this framework, the anomalous Hall conductivity was calculated for the $\pd$ Zener model\cite{dietl:2000} by Jungwirth \etal\cite{jungwirth:2002} The band structure of this model is composed of the six hole bands described by the Kohn-Luttinger Hamiltonian with the mean-field spin splitting, neglecting the spin-orbit induced Rashba (linear in $k$) and Dresselhaus ($k^3$) terms. While the former, together with all terms linear in $k$, does not generate the spin current,\cite{rashba} the latter does,\cite{malshukov:2005} which has not been hitherto studied in (Ga,Mn)As and related ferromagnets.

In this paper, we investigate the intrinsic anomalous Hall effect (for our purposes called the AHE) in the $\pd$ Zener model of a diluted ferromagnetic semiconductor, focusing on (Ga,Mn)As. This problem requires a complete description of the host band structure. We demonstrate it numerically by using different $\kp$ and tight-binding computational schemes described in Sec.~\ref{sec:thapproach}. We employ the 6-band\cite{kohn,dietl:2000, dietl:2001, abolfath:2001} and the 8-band $\kp$ model including the Dresselhaus splitting,\cite{bahder:1990,ostromek:1996} and two empirical tight-binding parameterizations\cite{jancu:1998,dicarlo:2003} ($spds^\star$ and $sps^\star$) to describe the GaAs band structure. The Mn$_\text{Ga}$ substitutions are introduced within the mean-field and virtual-crystal approximations. In Section~\ref{sec:results}, we first calibrate the models so as to obtain the agreement of two important characteristics of (Ga,Mn)As, the Curie temperature and uniaxial anisotropy. Then, we calculate the Berry curvature and the anomalous Hall conductivity to reveal qualitative differences between the calibrated models. We even report a negative conductivity sign within the new approaches, which was not observed in the previously employed 6-band $\kp$ model.\cite{jungwirth:2002} This result relates to the inversion asymmetry of the zinc-blende lattice, inherited by the Berry curvature. We provide the physical interpretation of our findings and briefly discuss their experimental implications.

\section{Theoretical approach}
\label{sec:thapproach}

We investigate AHE in a Hall sample of ferromagnetic (Ga,Mn)As with the electric field $\vec E \parallel \hat x$ and the magnetic field applied along the $[00\bar1]$ direction. This setup yields the anomalous conductivity
\begin{equation}
\label{eq:sigmaxy}
\sigma_{xy} = -\frac{e^2}{V \hbar} \langle \Omega_z \rangle\ ,
\end{equation}
where $\langle \Omega_z \rangle = \sum_{\vec{k}, n} \Omega_z(n,\vec{k}) f_{n,\vec{k}}$, and $f_{n,\vec{k}}$ is the Fermi-Dirac distribution associated with the band $n$ and wave vector $\vec{k}$. The positive values of $\sigma_{xy}$ mean that the anomalous Hall voltage has the same sign as in the ordinary Hall effect.

The Berry curvature in~\eqref{eq:sigmaxy} is given by
\begin{equation}
\label{eq:Omegaz-diff}
\Omega_z(n,\vec{k}) = 2 \im \langle \partial_{k_y} u_{n,\vec{k}} | \partial_{k_x} u_{n,\vec{k}} \rangle \ ,
\end{equation}
or by the equivalent Kubo formula (derived by differentiating the Schr\"{o}dinger equation over $\vec{k}$, which makes sense in our finite-dimensional Hi1bert space)
\begin{equation}
\label{eq:Omegaz-en}
\Omega_z(n,\vec{k}) = 2 \im \sum_{n' \neq n} \frac{c_{nn'}^y c_{n'n}^x}{(\epsilon_{n,\vec{k}} - \epsilon_{n',\vec{k}})^2} \ ,
\end{equation}
where $\vec c_{nn'} = \langle u_{n,\vec{k}} | \partial_{\vec k} \hat{H}_\vec{k} | u_{n',\vec{k}} \rangle$, and $u_{n,\vec{k}}$ are the periodic parts of the Bloch states with energies $\epsilon_{n,\vec{k}}$. Formula \eqref{eq:Omegaz-diff} may carry large error even when we describe the Bloch wave functions quite accurately, because it involves their derivatives. For instance, in the 6-band $\kp$ model,\cite{dietl:2001,abolfath:2001} we obtain an almost perfect description of the $p$-type bands around the $\Gamma$ point, but their derivatives in general include significant contributions from the states outside this space. On the other hand, the sum in~\eqref{eq:Omegaz-en} goes over all bands, not just the hole $p$-type ones. Even the detailed description of these bands only is, therefore, not sufficient to calculate the Berry curvature accurately. The model used must also have enough room to allow for the inversion symmetry breaking, an important property of GaAs lattice.\cite{dresselhaus:1955}

For the above reasons, we expect the multiband tight-binding models\cite{jancu:1998,dicarlo:2003} of the host band structure to be the most appropriate for the description of the Berry curvature. They automatically take into account the lack of inversion symmetry, as they distinguish Ga and As atoms. Contrary to the perturbative $\kp$ methods, they properly describe the Bloch states away from the center of the Brillouin zone, which makes them better suited to high hole concentrations. We use the $spds^\star$ Jancu\cite{jancu:1998} and $sps^\star$ di~Carlo\cite{dicarlo:2003} parameterizations, basing our numerical tight-binding implementation on the code by Strahberger \etal,\cite{strahberger:2000} employed previously in the studies of spin transport properties in modulated (Ga,Mn)As structures.\cite{Sankowski:2007a,Oszwaldowski:2006a} 

The impact of the inversion symmetry breaking on the Berry curvature is additionally tested in the 8-band $\kp$ model with the Dresselhaus term included, following Ostromek.\cite{ostromek:1996} He found that the magnitude of the Dresselhaus spin splitting of the conduction band in GaAs depends on the values of two poorly known parameters $A'$ and $B$ describing the spin-independent and spin-orbit related $\kp$ interaction of the conduction band with remote bands, respectively. Two sets of $A'$ and $B$ values reproduced the experimental magnitudes of spin splittings.\cite{ostromek:1996} We have adopted the set for which $A'=0$ instead of the alternative one with $A'=-14.7$\,eV\,\AA, which appears unrealistic.\cite{hermann:1977} For the chosen parametrization, at $B =0$ (inversion symmetry preserved) the remaining parameters correspond to the standard 6-band $\kp$ model: $\gamma_1=6.85$, $\gamma_2=2.9$, $\gamma_3=2.1$, and spin-orbit splitting $\Delta_\text{so} = 0.341$\,eV. The $k$-dependent part of the spin-orbit interaction has negligible effect on the investigated quantities, so we neglect it for clarity.

\begin{figure}[ht!]
\centering
\includegraphics[width=0.9 \linewidth]{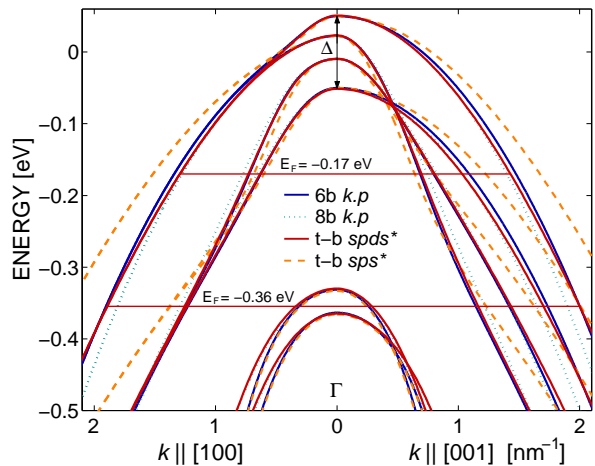}
\caption{[color online] The top of the GaAs valence band with spin splitting $\Delta=-0.1$ eV calculated by different methods. Horizontal lines denote the positions of the Fermi level $E_F$ for the hole densities 0.3 and 1.2 nm$^{-3}$ in the $spds^\star$ tight-binding model.}
\label{fig:bands}
\end{figure}

The dispersion relations of the top of the valence band calculated by the above methods are compared in Fig.\,\ref{fig:bands}. There is a very good agreement between the most popular 6-band $\kp$ and the most detailed $spds^\star$ tight-binding, as well as the 8-band $\kp$ model, while the results obtained within $sps^\star$ parameterization differ slightly.

The biaxial strain is included in particular computational schemes by adding an appropriate Bir-Pikus matrix to the 6-band\cite{dietl:2001,abolfath:2001} and to the 8-band $\kp$ model,\cite{bhusal:2004} or by changing the atoms arrangement in the tight-binding approaches, according to the strain tensor values: $\varepsilon_{xx} = \varepsilon_{yy} = \Delta a/a$ and $\varepsilon_{zz} = -2\, c_{12}/c_{11}\, \varepsilon_{xx}$, where $\Delta a$ is the strain-induced change of the lattice constant $a$, and $c_{12}/c_{11} = 0.453$ is the ratio of elastic moduli. Additionally, the on-site energies of the $d$ orbitals in the $spds^\star$ parameterization depend linearly on the strain tensor values.\cite{jancu:1998}

The AHE current flows in ferromagnetic (Ga,Mn)As, in which a part of Ga atoms is substituted by Mn ions producing strong $\pd$ hybridization. Each of them simultaneously forms a magnetic moment $S=5/2$ and creates one moderately bound hole. The holes fill the GaAs valence band, which is mostly of As $p$ character, from top to bottom. Their concentration $p$ is taken into account by adjusting the Fermi energy $E_F$. To do it efficiently, we assume that the crystal has a finite and very small volume $V\approx10^7 a_0^3$, and find the minimum of the first $n = pV$ occupied hole states' energies. The holes mediate ferromagnetic order between Mn spins $via$ exchange interactions. Within the mean-field and virtual crystal approximations, which we employ in all computational schemes, these interactions are spatially averaged. The constant molecular field of Mn spins creates a $k$-dependent Zeeman-like splitting of the host bands~(Fig.\,\ref{fig:bands}). This splitting for heavy holes in the $\Gamma$ point is given by $\Delta = x N_0 \langle S \rangle\beta$, where $x$ is the part of cation sites $N_0$ substituted by Mn with an average spin $\langle S \rangle$, and $\beta = -0.054$~eV\,nm$^3$ is the $\pd$ exchange integral.\cite{dietl:2001} Usually, the $\Delta$ value is smaller than it would result from the nominal Mn concentration. With increasing Mn doping, a part of Mn atoms occupies interstitial positions and tends to form pairs with substitutional ones, characterized by very small net magnetic moment.\cite{blinowski:2003} They can be removed by annealing, but the effective Mn concentration will remain smaller. Additionally, unintentional defects such as Mn interstitials and As antisites are double donors and reduce the hole concentration.


In both $\kp$ and tight-binding models, we have been able to compute the derivatives of the Hamiltonian matrix in~\eqref{eq:Omegaz-en} analytically, which significantly improves the accuracy of our results. Formula~\eqref{eq:Omegaz-diff} is equally suitable for numerical computation, if we overcome the problems created by the wavefunction phase gauge freedom, which is cancelled analytically, but not numerically. One should simply fix the phases of the relevant wave functions by dividing each of them by the phase factor of its first non-zero basis coefficient.

\section{Results}
\label{sec:results}

\subsection{Curie temperature and uniaxial anisotropy}

We begin with the comparison of the band structure models, looking at how they describe the two important characteristics of the $p$-type hole bands, the Curie temperature $T_{\mathrm{C}}$ and uniaxial anisotropy field $H_\text{un}$.

\begin{figure}[ht!]
\centering
\includegraphics[width=0.9 \linewidth]{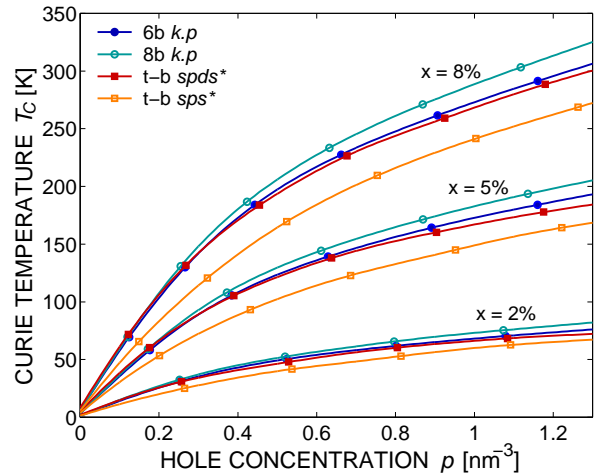}
\caption{[color online] Computed magnitudes of Curie temperatures as a function of the hole concentration according to various band structure models for Ga$_{1-x}$Mn$_x$As with the Mn content $x$ = 2\%, 5\% and 8\%. Vertical lines indicate the maximal experimentally realisable hole densities for given $x$. On this and the following graphs, the numerical data were generated more densely than the markers, which are visual aids only.}
\label{fig:pTc}
\end{figure}

The four models employed provide similar values of the Curie temperature $T_{\mathrm{C}}$, presented in Fig.\,\ref{fig:pTc} for three different Mn contents $x$, as a function of the hole concentration $p$. While the 6-band $\kp$ and $spds^\star$ tight-binding model give virtually identical $T_{\mathrm{C}}$ values, the remaining ones exhibit some differences due to their parameterization flaws (Fig.\,\ref{fig:bands}). The slight discrepancy between the two best results for high hole concentrations is resolved in favour of the more universal tight-binding approach. Since $T_{\mathrm{C}}$ is proportional to the thermodynamic spin density of states,\cite{dietl:2000,dietl:2001} we conclude that a mutually consistent description of the relevant valence bands is achieved throughout.

\begin{figure}[ht!]
\centering
\includegraphics[width=0.9 \linewidth]{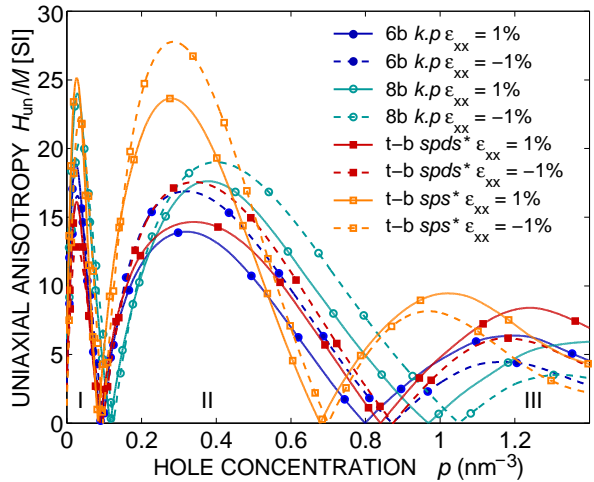}
\caption{[color online] The amplitude of the uniaxial anisotropy field $H_\text{un}$ (divided by saturation magnetization, see Ref.\,\onlinecite{dietl:2001}) for compressive ($\varepsilon_{xx} = -1\%$) and tensile ($\varepsilon_{xx} = 1\%$) biaxial strain in (Ga,Mn)As with spin splitting $\Delta = -0.15$\,eV. The effect of shape anisotropy is neglected. In the central range of hole concentrations (II), the easy axis is in-plane and perpendicular to the plane for compressive and tensile strain, respectively, while the reorientation transition is expected in either low or high concentration regimes (I or III).}
\label{fig:unax}
\end{figure}

A non-trivial comparison between the used models is provided by evaluating the magnitude of the uniaxial magnetic anisotropy field $H_\text{un}$ brought about by biaxial strain.\cite{dietl:2001,Zemen:2009a} This anisotropy is driven by the presence of the spin-orbit interaction in the carrier band. We calculate the magnitude of $H_\text{un}$ as proportional to the difference of the total carrier energy for the easy and hard magnetization directions under 1\% tensile or compressive strain (see Ref.\,\onlinecite{dietl:2001}). As presented in Fig.\,\ref{fig:unax}, in a region of intermediate hole concentrations, the easy axis takes the [001] direction for tensile strain, while it is in the (001) plane for compressive strain. The situation is opposite for lower and higher hole concentrations. These results agree between the models, especially for the $spds^\star$ tight-binding and the 6-band $\kp$ calculations. Consequently, all models handle similarly well the spin-orbit splitting of the $p$-type valence bands.

\subsection{Berry curvature and the related conductivity}

According to the previous chapter we can characterize both $T_{\mathrm{C}}$ and $H_\text{un}$ as \textit{static} quantities, which depend on the properties of the six occupied $p$-type bands only. Details of the other bands' structure, in particular the Dresselhaus $k^3$ splitting, do not influence their values. In marked contrast, the derivatives and interband elements in the Berry curvature formulas, \eqref{eq:Omegaz-diff} and \eqref{eq:Omegaz-en}, express the \textit{dynamic} character (related to the carrier drift caused by electric field) of the AHE and lead to qualitative differences between the models. Below we demonstrate the effect of the bulk inversion asymmetry on the Berry curvature and consequently on the anomalous conductivity trends.

\begin{figure}[ht!]
\centering
\includegraphics[width= \linewidth]{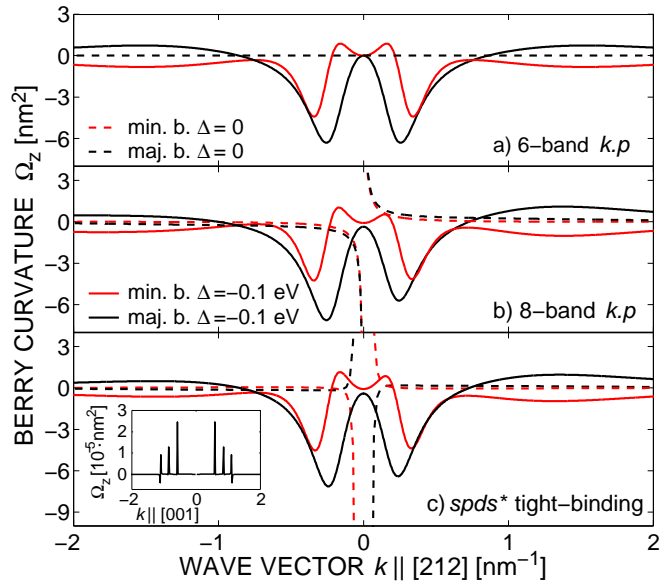}
\caption{[color online] Berry curvature of majority heavy and light hole bands in (Ga,Mn)As calculated using a) the 6-band $\kp$, b) 8-band $\kp$ and c) $spds^\star$ tight-binding model, with and without the spin splitting $\Delta$. Inset: Divergences in the Berry curvature.}
\label{fig:curvatures}
\end{figure}

The 6-band $\kp$ model describes the diamond lattice structure. Since the Kohn-Luttinger Hamiltonian it uses is invariant under space inversion, which is unitary, we have $\boldsymbol\Omega(-\vec k) = \boldsymbol\Omega(\vec k)$. On the other hand, the antiunitarity of time reversal operator leads to $\boldsymbol\Omega(-\vec k) = -\boldsymbol\Omega(\vec k)$ in the presence of the corresponding symmetry. Thus, the Berry curvature in this model is always symmetric and vanishes in the absence of magnetic fields, as presented in Fig.\,\ref{fig:curvatures}a, and no spin current flows.

The 8-band $\kp$ model contains, in addition, the $s$-type conduction band with the Dresselhaus spin splitting included by the use of L\"owdin perturbation calculus.\cite{bahder:1990,ostromek:1996} It results from the inversion symmetry breaking in the zinc-blende structure\cite{dresselhaus:1955} and thus, in the presence of the time-reversal symmetry, leads to non-vanishing antisymmetric Berry curvatures (Fig.\,\ref{fig:curvatures}b). Then, the related $k^3$ energy term in the conduction band spectrum accounts for a non-zero spin current within the intrinsic spin Hall effect.\cite{malshukov:2005} When the magnetic field is on, the significant asymmetry of the curvatures can still be observed.

The multiband tight-binding methods give us the detailed band parameterization and introduce realistic symmetries of the crystal lattice in a natural way. Figure~\ref{fig:curvatures}c presents the Berry curvatures obtained using the $spds^\star$ parameterization. Their symmetry is similar to the 8-band $k.p$ model, but the shape differs (especially for $\Delta=0$), pointing to the sensitivity of the spin topological effects to the subtleties of the band structure.

\begin{figure}[ht!]
\centering
\includegraphics[width=0.9 \linewidth]{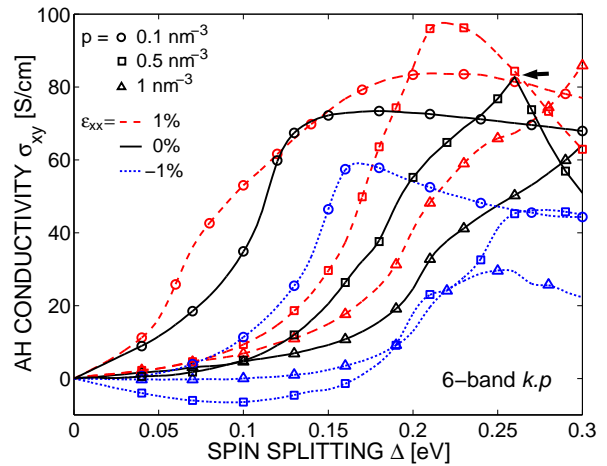}
\caption{[color online] Anomalous Hall conductivity $\sigma_{xy}$ \textit{vs} spin splitting $\Delta$ for different hole concentrations $p$ and biaxial strain $\varepsilon_{xx}$ in the 6-band $\kp$ model. The hallmark of the diabolic point for unstrained $p=0.5$\,nm$^{-3}$ curve is marked with an arrow. Numerical errors are ca 0.5\%.}
\label{fig:AHEIA_sxy6kp}
\end{figure}

An interesting effect is the formation of so-called diabolic points corresponding to the energy bands' crossings, best visible for $\Omega_z(\vec k)$ in the $\vec k \parallel [001]$ direction (Fig.\,\ref{fig:curvatures}c, inset). A commonly held view is that it is them which are the source of the anomalous Hall conductivity. Even though the degeneracies of states in the $k$-space do produce a nontrivial Berry potential, it is easy to show that their contributions to $\sigma_{xy}$ vanish for $T \to 0$~K. The coefficients $c^{x_i}_{nn'}$ in~\eqref{eq:Omegaz-en} are the matrix elements of Hermitian operators $\partial_{k_{x_i}} \hat{H}_\vec{k}$, hence $c^{x_i}_{nn'} = \cc{(c^{x_i}_{n'n})}$. The conductivity is thus proportional to the sum
\begin{equation}
\label{eq:contrib}
2 \im \sum_{\vec{k}} \sum_{n < n'} \frac{c_{nn'}^y c_{n'n}^x}{(\epsilon_{n,\vec{k}} - \epsilon_{n',\vec{k}})^2} \left( f_{n,\vec{k}}  - f_{n',\vec{k}} \right) \ .
\end{equation}
For $\epsilon_{n,\vec{k}} \neq \epsilon_{n',\vec{k}}$, a component of the above sum with given $(\vec{k},n,n')$ has a non-zero contribution to $\sigma_{xy}$ only if $f_{n,\vec{k}} \neq f_{n',\vec{k}}$, which for $T \rightarrow 0$ happens when one state is above and another below the Fermi level $E_F$. (Exploiting this observation in numerical computations ensures fast convergence of calculated integrals.) The component corresponding to the bands' crossing is thus zero, since $f_{n,\vec{k}} = f_{n',\vec{k}}$ in a neighbourhood of the diabolic point. For a diabolic point lying exactly at the Fermi level, the same follows from the fact that the crossing bands are always on the same side of $E_F$ in a neighbourhood of the diabolic point (which is always true for investigated systems, in which the Fermi level does not touch the borders of the Brillouin zone). Hence, for $T \to 0$ the diabolic points have no singular contribution to the anomalous Hall conductivity, which we also have confirmed numerically for finite temperatures. It is clearly seen from~\eqref{eq:Omegaz-diff} and~\eqref{eq:Omegaz-en} that the Berry curvature arises from the spin-orbit interaction. This is because Hamiltonians without the spin-orbit coupling operator have real representations for all $\vec{k}$. One can then choose $u_{n,\vec{k}}$ which are entirely real for all $\vec{k}$ and do not produce the Berry curvature. By the introduction of the spin-orbit coupling, the Hamiltonian becomes complex, causing the Berry curvature to arise. Yet, the diabolic points manifest themselves, when passing the Fermi level, as kinks in the conductivity (marked with an arrow in Fig.\,\ref{fig:AHEIA_sxy6kp}).

\begin{figure}[ht!]
\centering
\includegraphics[width=0.9 \linewidth]{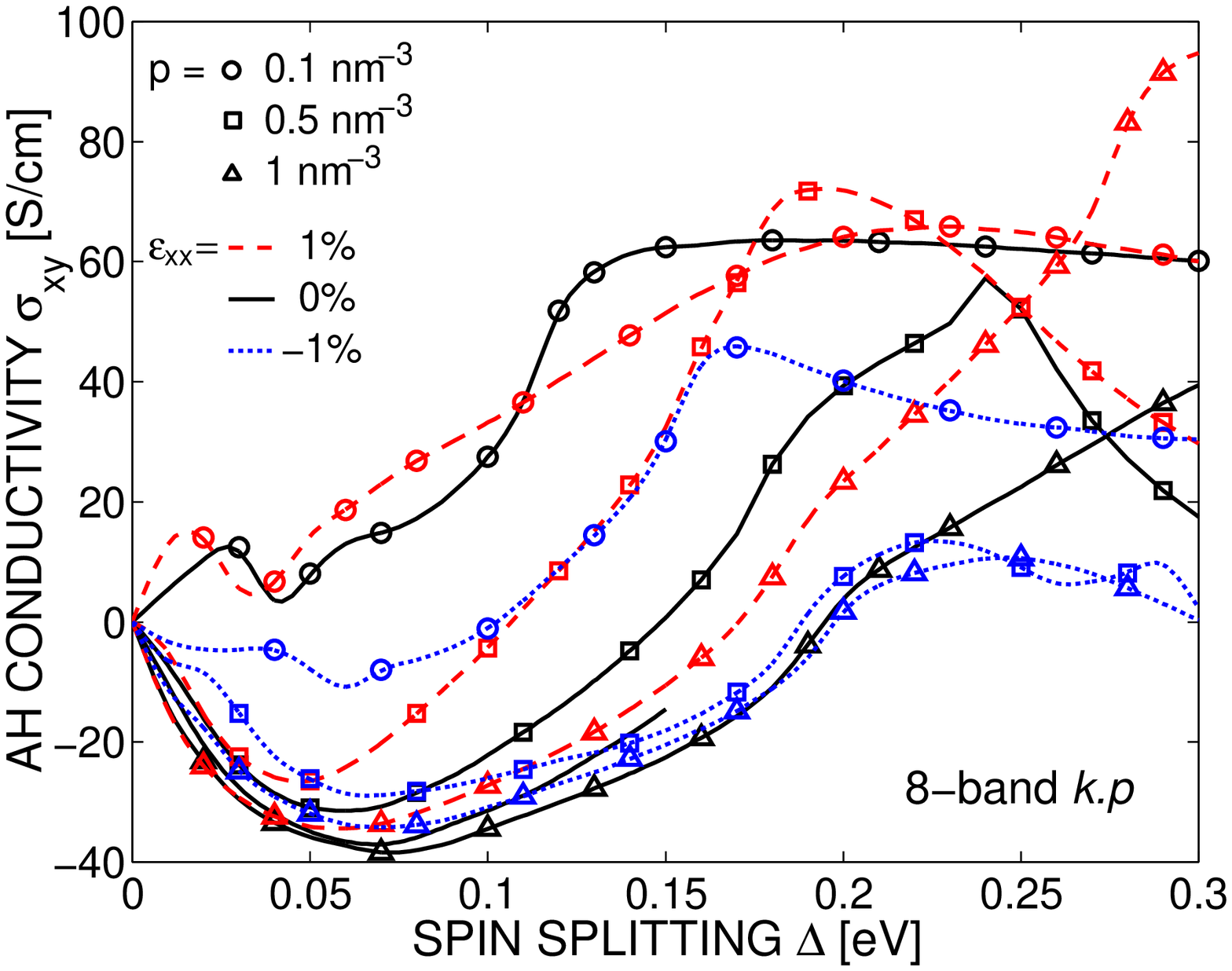}
\caption{[color online] Anomalous Hall conductivity $\sigma_{xy}$ \textit{vs} spin splitting $\Delta$ for different hole concentrations $p$ and biaxial strain $\varepsilon_{xx}$ in the 8-band $\kp$ model.}
\label{fig:AHEIA_sxy8kp}
\end{figure}

\begin{figure}[ht!]
\centering
\includegraphics[width=0.9 \linewidth]{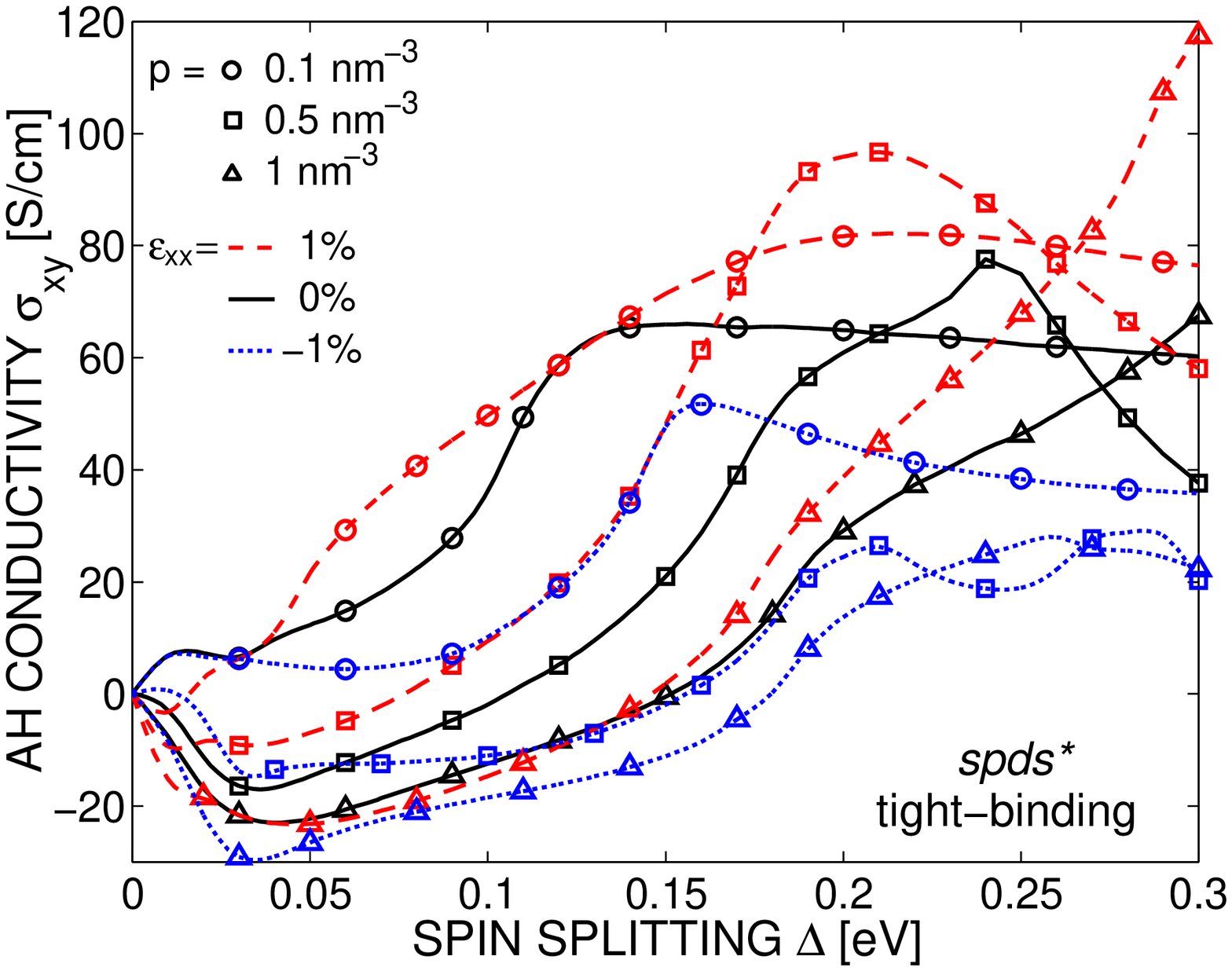}
\caption{[color online] Anomalous Hall conductivity $\sigma_{xy}$ \textit{vs} spin splitting $\Delta$ for different hole concentrations $p$ and biaxial strain $\varepsilon_{xx}$ in the $spds^\star$ tight-binding model.}
\label{fig:AHEIA_sxyJxy}
\end{figure}

\begin{figure}[ht!]
\centering
\includegraphics[width=0.9 \linewidth]{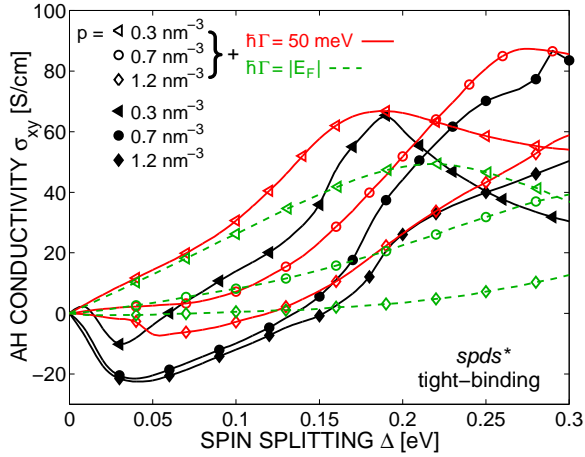}
\caption{[color online] Anomalous Hall conductivity $\sigma_{xy}$ \textit{vs} spin splitting $\Delta$ for different hole concentrations $p$ in the $spds^\star$ tight-binding model. The scattering-induced energy level broadening $\hbar \Gamma$ equal to 50\,meV and $|E_F|$ is included.}
\label{fig:AHEIA_sxyJxyhG}
\end{figure}

\begin{figure}[ht!]
\centering
\includegraphics[width=0.9 \linewidth]{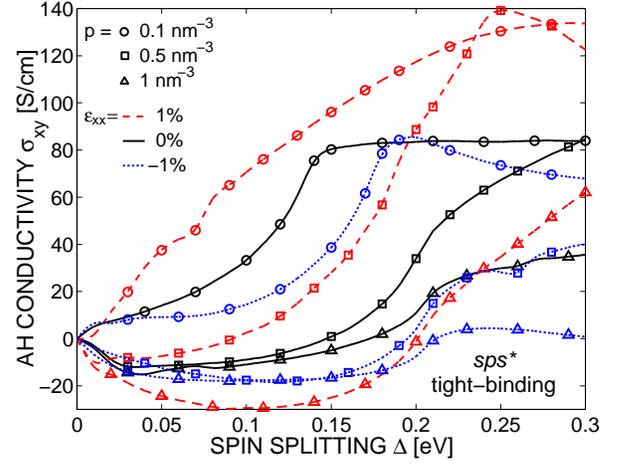}
\caption{[color online] Anomalous Hall conductivity \textit{vs} spin splitting for different hole concentrations $p$ and biaxial strain $\varepsilon_{xx}$ in the $sps^\star$ tight-binding model.}
\label{fig:AHEIA_sxydCxy}
\end{figure}

The qualitative difference of the Berry curvature between the models takes an effect on the anomalous Hall conductivity trends. The values of $\sigma_{xy}$ in the $\kp$ and tight-binding approaches, computed for various hole concentrations $p$ as a function of the valence band splitting $\Delta$, are presented in Figs\,\ref{fig:AHEIA_sxy6kp}-\ref{fig:AHEIA_sxydCxy}. The results obtained within the particular models remain in good agreement throughout the whole range of $\Delta$ values only for low hole concentrations, $p < 0.3$\,nm$^{-3}$. For higher carrier densities, differences in the $\sigma_{xy}$ values become significant, particularly for small and intermediate spin splittings. Remarkably, we obtain a negative sign of $\sigma_{xy}$ within the 8-band and tight-binding models (Figs\,\ref{fig:AHEIA_sxy8kp}-\ref{fig:AHEIA_sxydCxy}) in this range: the higher the hole concentration, the wider the range of $\Delta$ for which the negative sign persists. This is the effect of the Dresselhaus splitting which increases with $k$, while for increasing hole concentrations the states with high $k$-vectors become occupied and contribute to the conductivity. However, strong enough spin splitting destroys the negative sign. This shows a dramatic and so far unnoticed influence of the Dresselhaus term on the AHE in hole-controlled ferromagnetic semiconductors.

It has been suggested\cite{jungwirth:2003} that the influence of disorder on the intrinsic AHE can be phenomenologically modeled by substituting one of the energy differences in~\eqref{eq:Omegaz-en} with $\epsilon_{n,\vec{k}} - \epsilon_{n',\vec{k}} + i\hbar\Gamma$. The scattering-induced broadening of bands $\hbar\Gamma$ in (Ga,Mn)As at the localization boundary is presumably of the order of the Fermi energy $E_F$.\cite{jungwirth:2008} It washes out the Dresselhaus splitting and reduces the magnitude of its negative contribution to $\sigma_{xy}$, as shown in Fig.\,\ref{fig:AHEIA_sxyJxyhG}. However, this approach is not without its own problems: the energy level broadening is but a part of equal-rank ``extrinsic'' terms in the Kubo-St\v{r}eda formalism,\cite{sinitsyn:2008} and its magnitude is typically too large to treat its effect on the AHE perturbatively.

The sensitivity of the AHE to the details of the band structure suggests that it can be influenced by the biaxial strain. Figures~\ref{fig:AHEIA_sxy6kp}-\ref{fig:AHEIA_sxydCxy} contain the results on the anomalous Hall conductivity in tensile and compressively strained (Ga,Mn)As samples, $\varepsilon_{xx} = 1\%$ and $\varepsilon_{xx} = -1\%$. The $\sigma_{xy}$ values tend to increase in the first case and decrease in the latter in all models. As seen, small negative values are found already within the 6-band $\kp$ model for the tensile strain.

Additionally, we checked that despite the overall sensitivity, the effect of the temperature parameter in the Fermi-Dirac function on $\sigma_{xy}$ for a fixed value of spin splitting is negligibly small.

\subsection{Comparison to experiment}

\begin{figure}[ht!]
\centering
\includegraphics[width=\linewidth]{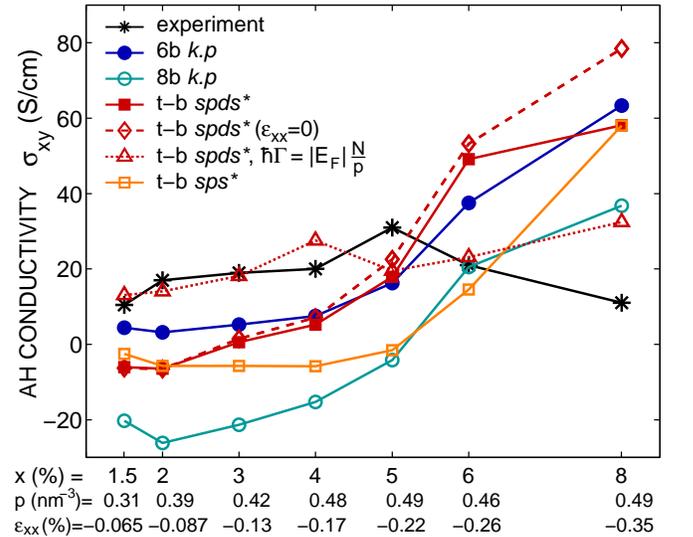}
\caption{[color online] Reconstruction of the experimental\cite{jungwirth:2003} anomalous Hall conductivities for nominal Mn concentration $x$, hole concentration $p$ and strain $\varepsilon_{xx}$, using different theoretical models.}
\label{fig:AHEIA_expsxy}
\end{figure}

Figure~\ref{fig:AHEIA_expsxy} compares the theoretical and experimental results\cite{jungwirth:2003, wang:2002} on the anomalous Hall conductivity for the set of annealed samples with nominal Mn concentration $x$, hole concentration $p$ and biaxial strain $\varepsilon_{xx}$. The calculations of the $\kp$ and tight-binding models corresponding to the experimental parameters do not fit the measured points. The new detailed theories, which include the inversion asymmetry of the GaAs lattice, predict a negative sign of $\sigma_{xy}$ for small Mn contents $x$. At the same time, all the models predict the $\sigma_{xy}$ values larger than in the experiment for high Mn content $x$. They are nevertheless significantly lowered by the---strong in this regime---biaxial strain, as shown by the comparison with zero strain calculations for $spds^\star$ tight-binding model (diamonds).

The much smaller than theoretical values of $\sigma_{xy}$ for samples with high Mn content may be connected to the presence of Mn interstitials,\cite{blinowski:2003} which do not form magnetic moments. The ones which survived the annealing process, as suggested by the measured hole densities, compensate one substitutional Mn spin each. As a result, the effective Mn concentration is lower than the total Mn content used in our calculations, which typically leads to lower $\sigma_{xy}$ values. This effect cannot explain the qualitative difference between the new theories and experiment at low Mn concentrations. However, some experimental data suggest that the negative $\sigma_{xy}$ can be found under conditions indicated by the present computations.\cite{ruzmetov:2004}

Additionally, the energy levels' lifetime broadening,\cite{jungwirth:2003} as a part of scattering effects derived within the Kubo-St\v{r}eda formalism, is taken into account in the $spds^\star$ model (triangles). The broadening $\hbar \Gamma$ is taken to be the ratio of the total Mn concentration $N = 4x/a_0^3$ and hole concentration $p$, times the magnitude of Fermi energy, $|E_F|$. As mentioned before (Fig.\,\ref{fig:AHEIA_sxyJxyhG} and related text), it is done in a rather phenomenological way, but nevertheless leads to much better agreement with the experimental data.

\section{Summary}

We have compared four models of the (Ga,Mn)As band structure with regards to their impact on Curie temperature, uniaxial anisotropy and the intrinsic anomalous Hall effect. We considered the 8-band $\kp$ and two tight-binding ($spds^\star$ and $sps^\star$) parameterizations, and compared their results with the previously employed 6-band $\kp$ approach. The first two quantities do not depend significantly on the model used, a consequence of their static nature. On the other hand, taking into account the details of the band structure beyond the six hole bands leads to qualitatively new results on the anomalous Hall effect, which is dynamic. In particular, the inversion asymmetry of the GaAs lattice described by the Dresselhaus $k^3$ term produces the negative anomalous conductivity sign. Despite using the more detailed models of the band structure, we have not obtained the agreement with the experiment---indeed, moved away from it. This is a symptom of the intrinsic AHE theory being insufficient to describe the observed phenomenon. Possible additional mechanisms which merit detailed investigation include scattering and localization. Their influence on the anomalous Hall effect in ferromagnetic semiconductors will certainly be the subject of future studies.

\begin{acknowledgments}
We thank D.\,Chiba, J.\,Majewski, F.\,Matsukura, H.\,Ohno, and R.\,Oszwa{\l}dowski for inspiring discussions. A.\,W. acknowledges support by the Scholarship of the President of Polish Academy of Sciences for doctoral students and T.\,D. acknowledges support from the European Research Council within the ``Ideas'' 7th Framework Programme of the EC (FunDMS Advanced Grant).
\end{acknowledgments}

\bibliography{AHE3D}

\begin{thebibliography}{44}
\expandafter\ifx\csname natexlab\endcsname\relax\def\natexlab#1{#1}\fi
\expandafter\ifx\csname bibnamefont\endcsname\relax
  \def\bibnamefont#1{#1}\fi
\expandafter\ifx\csname bibfnamefont\endcsname\relax
  \def\bibfnamefont#1{#1}\fi
\expandafter\ifx\csname citenamefont\endcsname\relax
  \def\citenamefont#1{#1}\fi
\expandafter\ifx\csname url\endcsname\relax
  \def\url#1{\texttt{#1}}\fi
\expandafter\ifx\csname urlprefix\endcsname\relax\def\urlprefix{URL }\fi
\providecommand{\bibinfo}[2]{#2}
\providecommand{\eprint}[2][]{\url{#2}}

\bibitem[{\citenamefont{Hall}(1880)}]{hall:1880}
\bibinfo{author}{\bibfnamefont{E.~H.} \bibnamefont{Hall}},
  \bibinfo{journal}{Philos. Mag.} \textbf{\bibinfo{volume}{10}},
  \bibinfo{pages}{301} (\bibinfo{year}{1880}).

\bibitem[{\citenamefont{Dyakonov}(2007)}]{dyakonov:2007}
\bibinfo{author}{\bibfnamefont{M.~I.} \bibnamefont{Dyakonov}},
  \bibinfo{journal}{Phys. Rev. Lett.} \textbf{\bibinfo{volume}{99}},
  \bibinfo{eid}{126601} (\bibinfo{year}{2007}).

\bibitem[{\citenamefont{Nagaosa et~al.}(2009)\citenamefont{Nagaosa, Sinova,
  Onoda, MacDonald, and Ong}}]{Nagaosa:2009a}
\bibinfo{author}{\bibfnamefont{N.}~\bibnamefont{Nagaosa}},
  \bibinfo{author}{\bibfnamefont{J.}~\bibnamefont{Sinova}},
  \bibinfo{author}{\bibfnamefont{S.}~\bibnamefont{Onoda}},
  \bibinfo{author}{\bibfnamefont{A.~H.} \bibnamefont{MacDonald}},
  \bibnamefont{and} \bibinfo{author}{\bibfnamefont{N.~P.} \bibnamefont{Ong}},
  \bibinfo{journal}{arXiv:0904.4154}  (\bibinfo{year}{2009}).

\bibitem[{\citenamefont{Smit}(1955)}]{smit:1955}
\bibinfo{author}{\bibfnamefont{J.}~\bibnamefont{Smit}},
  \bibinfo{journal}{Physica} \textbf{\bibinfo{volume}{21}},
  \bibinfo{pages}{877} (\bibinfo{year}{1955}).

\bibitem[{\citenamefont{Mott and Massey}(1965)}]{mott:1965}
\bibinfo{author}{\bibfnamefont{N.~F.} \bibnamefont{Mott}} \bibnamefont{and}
  \bibinfo{author}{\bibfnamefont{H.~S.~W.} \bibnamefont{Massey}},
  \emph{\bibinfo{title}{The Theory of Atomic Collisions}}
  (\bibinfo{publisher}{Clarendon Press}, \bibinfo{address}{Oxford},
  \bibinfo{year}{1965}), \bibinfo{edition}{3rd} ed.

\bibitem[{\citenamefont{Dyakonov and Perel}(1971)}]{dyakonov:1971}
\bibinfo{author}{\bibfnamefont{M.~I.} \bibnamefont{Dyakonov}} \bibnamefont{and}
  \bibinfo{author}{\bibfnamefont{V.~I.} \bibnamefont{Perel}},
  \bibinfo{journal}{Phys. Lett. A} \textbf{\bibinfo{volume}{35}},
  \bibinfo{pages}{459} (\bibinfo{year}{1971}).

\bibitem[{\citenamefont{Berger}(1970)}]{berger:1970}
\bibinfo{author}{\bibfnamefont{L.}~\bibnamefont{Berger}},
  \bibinfo{journal}{Phys. Rev. B} \textbf{\bibinfo{volume}{2}},
  \bibinfo{pages}{4559} (\bibinfo{year}{1970}).

\bibitem[{\citenamefont{Berger}(1972)}]{berger:1972}
\bibinfo{author}{\bibfnamefont{L.}~\bibnamefont{Berger}},
  \bibinfo{journal}{Phys. Rev. B} \textbf{\bibinfo{volume}{5}},
  \bibinfo{pages}{1862} (\bibinfo{year}{1972}).

\bibitem[{\citenamefont{Nozi\`{e}res and Lewiner}(1973)}]{nozieres:1973}
\bibinfo{author}{\bibfnamefont{P.}~\bibnamefont{Nozi\`{e}res}}
  \bibnamefont{and} \bibinfo{author}{\bibfnamefont{C.}~\bibnamefont{Lewiner}},
  \bibinfo{journal}{J. Phys. (Paris)} \textbf{\bibinfo{volume}{34}},
  \bibinfo{pages}{901} (\bibinfo{year}{1973}).

\bibitem[{\citenamefont{Dyakonov and Khaetskii}(1984)}]{dyakonov:1984}
\bibinfo{author}{\bibfnamefont{M.~I.} \bibnamefont{Dyakonov}} \bibnamefont{and}
  \bibinfo{author}{\bibfnamefont{A.~V.} \bibnamefont{Khaetskii}},
  \bibinfo{journal}{Z. Eksp. Teor. Fiz.} \textbf{\bibinfo{volume}{84}},
  \bibinfo{pages}{1843} (\bibinfo{year}{1984}).

\bibitem[{\citenamefont{Abanin et~al.}(2009)\citenamefont{Abanin, Shytov,
  Levitov, and Halperin}}]{abanin:2009}
\bibinfo{author}{\bibfnamefont{D.~A.} \bibnamefont{Abanin}},
  \bibinfo{author}{\bibfnamefont{A.~V.} \bibnamefont{Shytov}},
  \bibinfo{author}{\bibfnamefont{L.~S.} \bibnamefont{Levitov}},
  \bibnamefont{and} \bibinfo{author}{\bibfnamefont{B.~I.}
  \bibnamefont{Halperin}}, \bibinfo{journal}{Phys. Rev. B}
  \textbf{\bibinfo{volume}{79}}, \bibinfo{pages}{035304}
  (\bibinfo{year}{2009}).

\bibitem[{\citenamefont{Karplus and Luttinger}(1954)}]{karplus:1954}
\bibinfo{author}{\bibfnamefont{R.}~\bibnamefont{Karplus}} \bibnamefont{and}
  \bibinfo{author}{\bibfnamefont{J.~M.} \bibnamefont{Luttinger}},
  \bibinfo{journal}{Phys. Rev.} \textbf{\bibinfo{volume}{95}},
  \bibinfo{pages}{1154} (\bibinfo{year}{1954}).

\bibitem[{\citenamefont{Matl et~al.}(1998)\citenamefont{Matl, Ong, Yan, Li,
  Studebaker, Baum, and Doubinia}}]{matl:1998}
\bibinfo{author}{\bibfnamefont{P.}~\bibnamefont{Matl}},
  \bibinfo{author}{\bibfnamefont{N.~P.} \bibnamefont{Ong}},
  \bibinfo{author}{\bibfnamefont{Y.~F.} \bibnamefont{Yan}},
  \bibinfo{author}{\bibfnamefont{Y.~Q.} \bibnamefont{Li}},
  \bibinfo{author}{\bibfnamefont{D.}~\bibnamefont{Studebaker}},
  \bibinfo{author}{\bibfnamefont{T.}~\bibnamefont{Baum}}, \bibnamefont{and}
  \bibinfo{author}{\bibfnamefont{G.}~\bibnamefont{Doubinia}},
  \bibinfo{journal}{Phys. Rev. B} \textbf{\bibinfo{volume}{57}},
  \bibinfo{pages}{10248} (\bibinfo{year}{1998}).

\bibitem[{\citenamefont{Taguchi et~al.}(2001)\citenamefont{Taguchi, Oohara,
  Yoshizawa, Nagaosa, and Tokura}}]{taguchi:2001}
\bibinfo{author}{\bibfnamefont{Y.}~\bibnamefont{Taguchi}},
  \bibinfo{author}{\bibfnamefont{Y.}~\bibnamefont{Oohara}},
  \bibinfo{author}{\bibfnamefont{H.}~\bibnamefont{Yoshizawa}},
  \bibinfo{author}{\bibfnamefont{N.}~\bibnamefont{Nagaosa}}, \bibnamefont{and}
  \bibinfo{author}{\bibfnamefont{Y.}~\bibnamefont{Tokura}},
  \bibinfo{journal}{Science} \textbf{\bibinfo{volume}{291}},
  \bibinfo{pages}{2573} (\bibinfo{year}{2001}).

\bibitem[{\citenamefont{Jungwirth et~al.}(2002)\citenamefont{Jungwirth, Niu,
  and MacDonald}}]{jungwirth:2002}
\bibinfo{author}{\bibfnamefont{T.}~\bibnamefont{Jungwirth}},
  \bibinfo{author}{\bibfnamefont{Q.}~\bibnamefont{Niu}}, \bibnamefont{and}
  \bibinfo{author}{\bibfnamefont{A.~H.} \bibnamefont{MacDonald}},
  \bibinfo{journal}{Phys. Rev. Lett.} \textbf{\bibinfo{volume}{88}},
  \bibinfo{pages}{207208} (\bibinfo{year}{2002}).

\bibitem[{\citenamefont{Dietl et~al.}(2003)\citenamefont{Dietl, Matsukura,
  Ohno, Cibert, and Ferrand}}]{Dietl:2003c}
\bibinfo{author}{\bibfnamefont{T.}~\bibnamefont{Dietl}},
  \bibinfo{author}{\bibfnamefont{F.}~\bibnamefont{Matsukura}},
  \bibinfo{author}{\bibfnamefont{H.}~\bibnamefont{Ohno}},
  \bibinfo{author}{\bibfnamefont{J.}~\bibnamefont{Cibert}}, \bibnamefont{and}
  \bibinfo{author}{\bibfnamefont{D.}~\bibnamefont{Ferrand}}, in
  \emph{\bibinfo{booktitle}{Recent Trends in Theory of Physical Phenomena in
  High Magnetic Fields}}, edited by
  \bibinfo{editor}{\bibfnamefont{I.}~\bibnamefont{Vagner}}
  (\bibinfo{publisher}{Kluwer, Dordrecht}, \bibinfo{year}{2003}), p.
  \bibinfo{pages}{197}, \eprint{cond-mat/0306484}.

\bibitem[{\citenamefont{Lee et~al.}(2004{\natexlab{a}})\citenamefont{Lee,
  Watauchi, Miller, Cava, and Ong}}]{lee:2004a}
\bibinfo{author}{\bibfnamefont{W.-L.} \bibnamefont{Lee}},
  \bibinfo{author}{\bibfnamefont{S.}~\bibnamefont{Watauchi}},
  \bibinfo{author}{\bibfnamefont{V.~L.} \bibnamefont{Miller}},
  \bibinfo{author}{\bibfnamefont{R.~J.} \bibnamefont{Cava}}, \bibnamefont{and}
  \bibinfo{author}{\bibfnamefont{N.~P.} \bibnamefont{Ong}},
  \bibinfo{journal}{Science} \textbf{\bibinfo{volume}{303}},
  \bibinfo{pages}{1647} (\bibinfo{year}{2004}{\natexlab{a}}).

\bibitem[{\citenamefont{Lee et~al.}(2004{\natexlab{b}})\citenamefont{Lee,
  Watauchi, Miller, Cava, and Ong}}]{lee:2004b}
\bibinfo{author}{\bibfnamefont{W.-L.} \bibnamefont{Lee}},
  \bibinfo{author}{\bibfnamefont{S.}~\bibnamefont{Watauchi}},
  \bibinfo{author}{\bibfnamefont{V.~L.} \bibnamefont{Miller}},
  \bibinfo{author}{\bibfnamefont{R.~J.} \bibnamefont{Cava}}, \bibnamefont{and}
  \bibinfo{author}{\bibfnamefont{N.~P.} \bibnamefont{Ong}},
  \bibinfo{journal}{Phys. Rev. Lett.} \textbf{\bibinfo{volume}{93}},
  \bibinfo{pages}{226601} (\bibinfo{year}{2004}{\natexlab{b}}).

\bibitem[{\citenamefont{Berry}(1984)}]{berry:1984}
\bibinfo{author}{\bibfnamefont{M.~V.} \bibnamefont{Berry}},
  \bibinfo{journal}{Proc. R. Soc.} \textbf{\bibinfo{volume}{A 392}},
  \bibinfo{pages}{45} (\bibinfo{year}{1984}).

\bibitem[{\citenamefont{Sundaram and Niu}(1999)}]{sundaram:1999}
\bibinfo{author}{\bibfnamefont{G.}~\bibnamefont{Sundaram}} \bibnamefont{and}
  \bibinfo{author}{\bibfnamefont{Q.}~\bibnamefont{Niu}},
  \bibinfo{journal}{Phys. Rev. B} \textbf{\bibinfo{volume}{59}},
  \bibinfo{pages}{14915} (\bibinfo{year}{1999}).

\bibitem[{\citenamefont{Haldane}(2004)}]{haldane:2004}
\bibinfo{author}{\bibfnamefont{F.~D.~M.} \bibnamefont{Haldane}},
  \bibinfo{journal}{Phys. Rev. Lett.} \textbf{\bibinfo{volume}{93}},
  \bibinfo{pages}{206602} (\bibinfo{year}{2004}).

\bibitem[{\citenamefont{Sinitsyn}(2008)}]{sinitsyn:2008}
\bibinfo{author}{\bibfnamefont{N.~A.} \bibnamefont{Sinitsyn}},
  \bibinfo{journal}{J. Phys.: Cond. Matt.} \textbf{\bibinfo{volume}{20}},
  \bibinfo{pages}{023201} (\bibinfo{year}{2008}).

\bibitem[{\citenamefont{Dietl et~al.}(2000)\citenamefont{Dietl, Ohno,
  Matsukura, Cibert, and Ferrand}}]{dietl:2000}
\bibinfo{author}{\bibfnamefont{T.}~\bibnamefont{Dietl}},
  \bibinfo{author}{\bibfnamefont{H.}~\bibnamefont{Ohno}},
  \bibinfo{author}{\bibfnamefont{F.}~\bibnamefont{Matsukura}},
  \bibinfo{author}{\bibfnamefont{J.}~\bibnamefont{Cibert}}, \bibnamefont{and}
  \bibinfo{author}{\bibfnamefont{D.}~\bibnamefont{Ferrand}},
  \bibinfo{journal}{Science} \textbf{\bibinfo{volume}{287}},
  \bibinfo{pages}{1019} (\bibinfo{year}{2000}).

\bibitem[{\citenamefont{Rashba}(2004)}]{rashba}
\bibinfo{author}{\bibfnamefont{E.~I.} \bibnamefont{Rashba}},
  \bibinfo{journal}{Phys. Rev. B} \textbf{\bibinfo{volume}{70}},
  \bibinfo{pages}{201309} (\bibinfo{year}{2004}).

\bibitem[{\citenamefont{Mal'shukov and Chao}(2005)}]{malshukov:2005}
\bibinfo{author}{\bibfnamefont{A.~G.} \bibnamefont{Mal'shukov}}
  \bibnamefont{and} \bibinfo{author}{\bibfnamefont{K.~A.} \bibnamefont{Chao}},
  \bibinfo{journal}{Phys. Rev. B} \textbf{\bibinfo{volume}{71}},
  \bibinfo{pages}{121308} (\bibinfo{year}{2005}).

\bibitem[{\citenamefont{Luttinger and Kohn}(1955)}]{kohn}
\bibinfo{author}{\bibfnamefont{J.~M.} \bibnamefont{Luttinger}}
  \bibnamefont{and} \bibinfo{author}{\bibfnamefont{W.}~\bibnamefont{Kohn}},
  \bibinfo{journal}{Phys. Rev.} \textbf{\bibinfo{volume}{97}},
  \bibinfo{pages}{869} (\bibinfo{year}{1955}).

\bibitem[{\citenamefont{Dietl et~al.}(2001)\citenamefont{Dietl, Ohno, and
  Matsukura}}]{dietl:2001}
\bibinfo{author}{\bibfnamefont{T.}~\bibnamefont{Dietl}},
  \bibinfo{author}{\bibfnamefont{H.}~\bibnamefont{Ohno}}, \bibnamefont{and}
  \bibinfo{author}{\bibfnamefont{F.}~\bibnamefont{Matsukura}},
  \bibinfo{journal}{Phys. Rev. B} \textbf{\bibinfo{volume}{63}},
  \bibinfo{pages}{195205} (\bibinfo{year}{2001}).

\bibitem[{\citenamefont{Abolfath et~al.}(2001)\citenamefont{Abolfath,
  Jungwirth, Brum, and MacDonald}}]{abolfath:2001}
\bibinfo{author}{\bibfnamefont{M.}~\bibnamefont{Abolfath}},
  \bibinfo{author}{\bibfnamefont{T.}~\bibnamefont{Jungwirth}},
  \bibinfo{author}{\bibfnamefont{J.}~\bibnamefont{Brum}}, \bibnamefont{and}
  \bibinfo{author}{\bibfnamefont{A.~H.} \bibnamefont{MacDonald}},
  \bibinfo{journal}{Phys. Rev. B} \textbf{\bibinfo{volume}{63}},
  \bibinfo{pages}{054418} (\bibinfo{year}{2001}).

\bibitem[{\citenamefont{Bahder}(1990)}]{bahder:1990}
\bibinfo{author}{\bibfnamefont{T.~B.} \bibnamefont{Bahder}},
  \bibinfo{journal}{Phys. Rev. B} \textbf{\bibinfo{volume}{41}},
  \bibinfo{pages}{11992} (\bibinfo{year}{1990}).

\bibitem[{\citenamefont{Ostromek}(1996)}]{ostromek:1996}
\bibinfo{author}{\bibfnamefont{T.~E.} \bibnamefont{Ostromek}},
  \bibinfo{journal}{Phys. Rev. B} \textbf{\bibinfo{volume}{54}},
  \bibinfo{pages}{14467} (\bibinfo{year}{1996}).

\bibitem[{\citenamefont{Jancu et~al.}(1998)\citenamefont{Jancu, Scholz,
  Beltram, and Bassani}}]{jancu:1998}
\bibinfo{author}{\bibfnamefont{J.-M.} \bibnamefont{Jancu}},
  \bibinfo{author}{\bibfnamefont{R.}~\bibnamefont{Scholz}},
  \bibinfo{author}{\bibfnamefont{F.}~\bibnamefont{Beltram}}, \bibnamefont{and}
  \bibinfo{author}{\bibfnamefont{F.}~\bibnamefont{Bassani}},
  \bibinfo{journal}{Phys. Rev. B} \textbf{\bibinfo{volume}{57}},
  \bibinfo{pages}{6493} (\bibinfo{year}{1998}).

\bibitem[{\citenamefont{Di~Carlo}(2003)}]{dicarlo:2003}
\bibinfo{author}{\bibfnamefont{A.}~\bibnamefont{Di~Carlo}},
  \bibinfo{journal}{Semicond. Sci. Technol.} \textbf{\bibinfo{volume}{18}},
  \bibinfo{pages}{R1} (\bibinfo{year}{2003}).

\bibitem[{\citenamefont{Dresselhaus}(1955)}]{dresselhaus:1955}
\bibinfo{author}{\bibfnamefont{G.}~\bibnamefont{Dresselhaus}},
  \bibinfo{journal}{Phys. Rev.} \textbf{\bibinfo{volume}{100}},
  \bibinfo{pages}{580} (\bibinfo{year}{1955}).

\bibitem[{\citenamefont{Strahberger and Vogl}(2000)}]{strahberger:2000}
\bibinfo{author}{\bibfnamefont{C.}~\bibnamefont{Strahberger}} \bibnamefont{and}
  \bibinfo{author}{\bibfnamefont{P.}~\bibnamefont{Vogl}},
  \bibinfo{journal}{Phys. Rev. B} \textbf{\bibinfo{volume}{62}},
  \bibinfo{pages}{7289} (\bibinfo{year}{2000}).

\bibitem[{\citenamefont{Sankowski et~al.}(2007)\citenamefont{Sankowski, Kacman,
  Majewski, and Dietl}}]{Sankowski:2007a}
\bibinfo{author}{\bibfnamefont{P.}~\bibnamefont{Sankowski}},
  \bibinfo{author}{\bibfnamefont{P.}~\bibnamefont{Kacman}},
  \bibinfo{author}{\bibfnamefont{J.~A.} \bibnamefont{Majewski}},
  \bibnamefont{and} \bibinfo{author}{\bibfnamefont{T.}~\bibnamefont{Dietl}},
  \bibinfo{journal}{Phys. Rev. B} \textbf{\bibinfo{volume}{75}},
  \bibinfo{eid}{045306} (\bibinfo{year}{2007}).

\bibitem[{\citenamefont{Oszwa{\l}dowski
  et~al.}(2006)\citenamefont{Oszwa{\l}dowski, Majewski, and
  Dietl}}]{Oszwaldowski:2006a}
\bibinfo{author}{\bibfnamefont{R.}~\bibnamefont{Oszwa{\l}dowski}},
  \bibinfo{author}{\bibfnamefont{J.~A.} \bibnamefont{Majewski}},
  \bibnamefont{and} \bibinfo{author}{\bibfnamefont{T.}~\bibnamefont{Dietl}},
  \bibinfo{journal}{Phys. Rev. B} \textbf{\bibinfo{volume}{74}},
  \bibinfo{pages}{153310} (\bibinfo{year}{2006}).

\bibitem[{\citenamefont{Hermann and Weisbuch}(1977)}]{hermann:1977}
\bibinfo{author}{\bibfnamefont{C.}~\bibnamefont{Hermann}} \bibnamefont{and}
  \bibinfo{author}{\bibfnamefont{C.}~\bibnamefont{Weisbuch}},
  \bibinfo{journal}{Phys. Rev. B} \textbf{\bibinfo{volume}{15}},
  \bibinfo{pages}{823} (\bibinfo{year}{1977}).

\bibitem[{\citenamefont{Bhusal et~al.}(2004)\citenamefont{Bhusal, Alemu, and
  Freundlich}}]{bhusal:2004}
\bibinfo{author}{\bibfnamefont{L.}~\bibnamefont{Bhusal}},
  \bibinfo{author}{\bibfnamefont{A.}~\bibnamefont{Alemu}}, \bibnamefont{and}
  \bibinfo{author}{\bibfnamefont{A.}~\bibnamefont{Freundlich}},
  \bibinfo{journal}{Nanotechnology} \textbf{\bibinfo{volume}{15}},
  \bibinfo{pages}{S245} (\bibinfo{year}{2004}).

\bibitem[{\citenamefont{Blinowski and Kacman}(2003)}]{blinowski:2003}
\bibinfo{author}{\bibfnamefont{J.}~\bibnamefont{Blinowski}} \bibnamefont{and}
  \bibinfo{author}{\bibfnamefont{P.}~\bibnamefont{Kacman}},
  \bibinfo{journal}{Phys. Rev. B} \textbf{\bibinfo{volume}{67}},
  \bibinfo{pages}{121204} (\bibinfo{year}{2003}).

\bibitem[{\citenamefont{Zemen et~al.}(2009)\citenamefont{Zemen,
  Ku\ifmmode~\check{c}\else \v{c}\fi{}era, Olejn\'ik, and
  Jungwirth}}]{Zemen:2009a}
\bibinfo{author}{\bibfnamefont{J.}~\bibnamefont{Zemen}},
  \bibinfo{author}{\bibfnamefont{J.}~\bibnamefont{Ku\ifmmode~\check{c}\else
  \v{c}\fi{}era}}, \bibinfo{author}{\bibfnamefont{K.}~\bibnamefont{Olejn\'ik}},
  \bibnamefont{and}
  \bibinfo{author}{\bibfnamefont{T.}~\bibnamefont{Jungwirth}},
  \bibinfo{journal}{Phys. Rev. B} \textbf{\bibinfo{volume}{80}},
  \bibinfo{pages}{155203} (\bibinfo{year}{2009}).

\bibitem[{\citenamefont{Jungwirth et~al.}(2003)\citenamefont{Jungwirth, Sinova,
  Wang, Edmonds, Campion, Gallagher, Foxon, Niu, and
  MacDonald}}]{jungwirth:2003}
\bibinfo{author}{\bibfnamefont{T.}~\bibnamefont{Jungwirth}},
  \bibinfo{author}{\bibfnamefont{J.}~\bibnamefont{Sinova}},
  \bibinfo{author}{\bibfnamefont{K.~Y.} \bibnamefont{Wang}},
  \bibinfo{author}{\bibfnamefont{K.~W.} \bibnamefont{Edmonds}},
  \bibinfo{author}{\bibfnamefont{R.~P.} \bibnamefont{Campion}},
  \bibinfo{author}{\bibfnamefont{B.~L.} \bibnamefont{Gallagher}},
  \bibinfo{author}{\bibfnamefont{C.~T.} \bibnamefont{Foxon}},
  \bibinfo{author}{\bibfnamefont{Q.}~\bibnamefont{Niu}}, \bibnamefont{and}
  \bibinfo{author}{\bibfnamefont{A.~H.} \bibnamefont{MacDonald}},
  \bibinfo{journal}{Appl. Phys. Lett.} \textbf{\bibinfo{volume}{83}},
  \bibinfo{pages}{320} (\bibinfo{year}{2003}).

\bibitem[{\citenamefont{Jungwirth et~al.}(2008)\citenamefont{Jungwirth,
  Gallagher, and Wunderlich}}]{jungwirth:2008}
\bibinfo{author}{\bibfnamefont{T.}~\bibnamefont{Jungwirth}},
  \bibinfo{author}{\bibfnamefont{B.~L.} \bibnamefont{Gallagher}},
  \bibnamefont{and}
  \bibinfo{author}{\bibfnamefont{J.}~\bibnamefont{Wunderlich}}, in
  \emph{\bibinfo{booktitle}{Spintroncs, Semiconductors and Semimetals}}, edited
  by \bibinfo{editor}{\bibfnamefont{T.}~\bibnamefont{Dietl}}
  (\bibinfo{publisher}{Elsevier}, \bibinfo{address}{Amsterdam},
  \bibinfo{year}{2008}), vol.~\bibinfo{volume}{82}, pp.
  \bibinfo{pages}{135--205}.

\bibitem[{\citenamefont{Wang et~al.}(2002)\citenamefont{Wang, Edmonds, Campion,
  Zhao, Neumann, Foxon, Gallagher, and Main}}]{wang:2002}
\bibinfo{author}{\bibfnamefont{K.~Y.} \bibnamefont{Wang}},
  \bibinfo{author}{\bibfnamefont{K.~W.} \bibnamefont{Edmonds}},
  \bibinfo{author}{\bibfnamefont{R.~P.} \bibnamefont{Campion}},
  \bibinfo{author}{\bibfnamefont{L.~X.} \bibnamefont{Zhao}},
  \bibinfo{author}{\bibfnamefont{A.~C.} \bibnamefont{Neumann}},
  \bibinfo{author}{\bibfnamefont{C.~T.} \bibnamefont{Foxon}},
  \bibinfo{author}{\bibfnamefont{B.~L.} \bibnamefont{Gallagher}},
  \bibnamefont{and} \bibinfo{author}{\bibfnamefont{P.~C.} \bibnamefont{Main}},
  \bibinfo{journal}{arXiv:cond-mat/0211697}  (\bibinfo{year}{2002}).

\bibitem[{\citenamefont{Ruzmetov et~al.}(2004)\citenamefont{Ruzmetov,
  Scherschligt, Baxter, Wojtowicz, Liu, Sasaki, Furdyna, Yu, and
  Walukiewicz}}]{ruzmetov:2004}
\bibinfo{author}{\bibfnamefont{D.}~\bibnamefont{Ruzmetov}},
  \bibinfo{author}{\bibfnamefont{J.}~\bibnamefont{Scherschligt}},
  \bibinfo{author}{\bibfnamefont{D.}~\bibnamefont{Baxter}},
  \bibinfo{author}{\bibfnamefont{T.}~\bibnamefont{Wojtowicz}},
  \bibinfo{author}{\bibfnamefont{X.}~\bibnamefont{Liu}},
  \bibinfo{author}{\bibfnamefont{Y.}~\bibnamefont{Sasaki}},
  \bibinfo{author}{\bibfnamefont{J.~K.} \bibnamefont{Furdyna}},
  \bibinfo{author}{\bibfnamefont{K.~M.} \bibnamefont{Yu}}, \bibnamefont{and}
  \bibinfo{author}{\bibfnamefont{W.}~\bibnamefont{Walukiewicz}},
  \bibinfo{journal}{Phys. Rev. B} \textbf{\bibinfo{volume}{69}},
  \bibinfo{pages}{155207} (\bibinfo{year}{2004}).

\end{thebibliography}

\end{document}